\documentclass[sigconf, nonacm, screen]{acmart}

\PassOptionsToPackage{prologue,table}{xcolor}
\PassOptionsToPackage{hyphens}{url}

\usepackage{xcolor}
\usepackage{wasysym}
\usepackage{framed}
\usepackage{booktabs}
\usepackage{ifthen}
\usepackage{color}
\usepackage{url}
\usepackage{xspace}
\usepackage[T1]{fontenc}
\usepackage{enumerate}
\usepackage[linesnumbered,ruled,vlined]{algorithm2e}
\usepackage{array,multirow,graphicx}
\usepackage{float}
\usepackage{balance}
\usepackage{tikz}
\usepackage{calc}
\usepackage{subfigure}
\usepackage{amsfonts}
\usepackage{pifont}
\usepackage{xspace}
\usepackage{hyperref,endnotes}
\usepackage{makecell}
\usepackage[misc]{ifsym}
\usepackage[noend]{algpseudocode}
\usepackage{booktabs}
\usepackage{algorithmicx}
\usepackage[table]{xcolor}
\usepackage{longtable}
\usepackage{listings}

\lstdefinelanguage{Solidity}{
	keywords=[1]{anonymous, assembly, assert, balance, break, call, callcode, case, catch, class, constant, continue, constructor, contract, debugger, default, delegatecall, delete, do, else, emit, event, experimental, export, external, false, finally, for, function, gas, if, implements, import, in, indexed, instanceof, interface, internal, is, length, library, log0, log1, log2, log3, log4, memory, modifier, new, payable, pragma, private, protected, public, pure, push, require, return, returns, revert, selfdestruct, send, solidity, storage, struct, suicide, super, switch, then, this, throw, transfer, true, try, typeof, using, value, view, while, with, addmod, ecrecover, keccak256, mulmod, ripemd160, sha256, sha3}, 
	keywordstyle=[1]\color{blue}\bfseries,
	keywords=[2]{address, bool, byte, bytes, bytes1, bytes2, bytes3, bytes4, bytes5, bytes6, bytes7, bytes8, bytes9, bytes10, bytes11, bytes12, bytes13, bytes14, bytes15, bytes16, bytes17, bytes18, bytes19, bytes20, bytes21, bytes22, bytes23, bytes24, bytes25, bytes26, bytes27, bytes28, bytes29, bytes30, bytes31, bytes32, enum, int, int8, int16, int24, int32, int40, int48, int56, int64, int72, int80, int88, int96, int104, int112, int120, int128, int136, int144, int152, int160, int168, int176, int184, int192, int200, int208, int216, int224, int232, int240, int248, int256, mapping, string, uint, uint8, uint16, uint24, uint32, uint40, uint48, uint56, uint64, uint72, uint80, uint88, uint96, uint104, uint112, uint120, uint128, uint136, uint144, uint152, uint160, uint168, uint176, uint184, uint192, uint200, uint208, uint216, uint224, uint232, uint240, uint248, uint256, var, void, ether, finney, szabo, wei, days, hours, minutes, seconds, weeks, years},	
	keywordstyle=[2]\color{teal}\bfseries,
	keywords=[3]{block, blockhash, coinbase, difficulty, gaslimit, number, timestamp, msg, data, gas, sender, sig, value, now, tx, gasprice, origin},	
	keywordstyle=[3]\color{violet}\bfseries,
	identifierstyle=\color{black},
	sensitive=false,
	comment=[l]{//},
	morecomment=[s]{/*}{*/},
	commentstyle=\color{gray}\ttfamily,
	stringstyle=\color{red}\ttfamily,
	morestring=[b]',
	morestring=[b]"
}

\lstdefinestyle{solidity}{
	language=Solidity,
	extendedchars=true,
	basicstyle=\footnotesize\ttfamily,
	showstringspaces=false,
	showspaces=false,
	numbers=left,
	numberstyle=\footnotesize,
	numbersep=9pt,
	tabsize=2,
	breaklines=true,
	showtabs=false,
	captionpos=b,
        float=t
}

\def\etal{\emph{et al.}\xspace}
\def\vs{\emph{vs. }\xspace}

\begin{document}

\title{SoK: On the Security of Non-Fungible Tokens}
\author{Kai Ma}
\affiliation{
  \institution{Huazhong University of Science and Technology}
  \country{China}
}
\author{Jintao Huang}
\affiliation{
  \institution{Huazhong University of Science and Technology}
  \country{China}
}

\author{Ningyu He}
\affiliation{
  \institution{Peking University}
  \country{China}
}

\author{Zhuo Wang}
\affiliation{
  \institution{Huazhong University of Science and Technology}
  \country{China}
}
\author{Haoyu Wang}
\affiliation{
  \institution{Huazhong University of Science and Technology}
  \country{China}
}

\thanks{*~The first three authors contributed equally to this work.}

\begin{abstract}

Non-fungible tokens (NFTs) drive the prosperity of the Web3 ecosystem. By November 2023, the total market value of NFT projects reached approximately 16 billion USD. Accompanying the success of NFTs are various security issues, i.e., attacks and scams are prevalent in the ecosystem.
While NFTs have attracted significant attentions from both industry and academia, there is a lack of understanding of kinds of NFT security issues.
The discovery, in-depth analysis, and systematic categorization of these security issues are of significant importance for the prosperous development of the NFT ecosystem. 
To fill the gap, we performed a systematic literature review related to NFT security, and we have identified 142 incidents from 213 security reports and 18 academic papers until October 1st, 2023. 
Through manual analysis of the compiled security incidents, we have classified them into 12 major categories. 
Then we explored potential solutions and mitigation strategies. Drawing from these analyses, we established the first NFT security reference frame.
Except, we extracted the characteristics of NFT security issues, i.e., the prevalence, severity, and intractability. We have indicated the gap between industry and academy for NFT security, and provide further research directions for the community.
This paper, as the first SoK of NFT security, has systematically explored the security issues within the NFT ecosystem, shedding light on their root causes, real-world attacks, and potential ways to address them. 
Our findings will contribute to the future research of NFT security.

\end{abstract}

\maketitle

\section{Introduction}

Since the beginning of 2021, non-fungible tokens (NFTs), a groundbreaking form of currency, have surged in popularity~\cite{nftpopularity}.
The most notable feature of NFT is its \textit{uniqueness}. In other words, there is a cryptocurrency-based algorithm to guarantee that it is impossible to counterfeit an existing NFT.
Taking advantage of this characteristic, as well as the anonymity and traceability of blockchain techniques, users can bind an NFT to a real-world asset, such as images, videos, or in-game items, through a URI (Uniform Resource Identifier)~\cite{tokenuri}.
To this end, NFT has greatly reduced the transaction costs of goods, and many NFT projects have received considerable investment as a result. For example, one NFT project called \textit{BAYC}~\cite{BAYC} has reached a total volume of over 1.4M Ether~\cite{BAYCvolume}.

However, alongside their increasing popularity, the whole NFT ecosystem is under numerous security threats that have resulted in millions of dollars in losses for investors and project owners.
For example, in April 2022, an individual lost his Bored Ape Yacht Club (BAYC) token worth over \$3M in a fraudulent swap transaction~\cite{Swapissue}. 
Similarly, in May 2022, a family was scammed out of \$1.4M by a perpetrator who sold them a counterfeit Moonbirds token~\cite{mayhack}, which had taken them years to accumulate. 
In September 2022, hackers stole Ethereum valued at \$185K from Bill Murray after he auctioned an NFT for charity~\cite{sephack}.
According to a research report~\cite{attacks}, there are currently more than 200 attack events against the NFT ecosystem, causing a total of more than \$88M in economic losses. More importantly, the frequency of these attacks continues to increase year by year.

In recent years, some studies about NFT security have been conducted. For example, Das et al.~\cite{das2022understanding} have conducted the first systematic analysis of NFT security.
Stoger et al. ~\cite{stoger2023demystifying}  found that almost every interaction of a user with a centralized entity can be exploited to hijack NFTs or cryptocurrencies from the user.
However, they only focus on the NFT market manipulation in the corresponding NFT projects, without paying attention on other components of the NFT ecosystem, like NFT smart contracts, which have also been exploited by adversaries to gain illegal profits.

To fill this gap, to the best of our knowledge, \textit{we have conducted the first SoK on the security side of the whole NFT ecosystem}. Against the NFT ecosystem, our goal is to summarize all happened attack events, uncover all their root causes, and propose the corresponding best practices for developers and investors.

\textbf{This work.}
In this SoK, we first constructed an NFT security reference frame in \S\ref{sec:reference:frame}. We collected the data from both security reports and academic papers and then extracted the incidents and security issues. 
Next, based on the reference frame, we dive deep into the contract layer in \S\ref{sec:contract:layer}, the market layer in \S\ref{sec:market:layer}, the auxiliary service layer in \S\ref{sec:auxiliary:service}, separately.
Specifically, we further discuss the characteristics, current research, and potential detection/defense methods of the extracted security issues.
We give our key findings and provide the discussion as well as the potential future research direction in \S\ref{sec:discussion}.

Our contributions and key findings are as follows.
\begin{itemize}
    \item \textbf{\textit{To the best of our knowledge, we have conducted the first SoK on NFT security, }}encompassing both prominent and lesser-known security issues.
    Specifically, we have compiled the first NFT security reference frame, with 142 incidents belonging to 12 major categories, which will serve as a general model for NFT security.

    \item \textbf{\textit{We have extracted the characteristics of NFT security issues.} }
    We demystify these NFT security issues from characteristics including the prevalence, severity, and intractability. 
    These characteristics serve as a compelling motivation to delve into the study of these security issues.

    \item \textbf{\textit{We have pointed out the huge gap between industry and academia.}}
    Our result shows that current research on NFT security only addresses a limited portion of NFT security issues that have been discovered by the industry.

\end{itemize}

We have released our crafted dataset at \href{https://nftsok.github.io/}{link}.

\section{BACKGROUND}
\label{sec:nft:background}
\subsection{Blockchain \& Cryptocurrency}
\label{sec:nft:background:primer}
Blockchain implements a distributed ledger that utilizes cryptography to securely link growing lists of records, i.e. \textit{blocks}~\cite{blockinblockchain}, together. 
Each block contains a set of transactions generated by \textit{accounts}~\cite{ethreumaccount}, i.e., the participants in the blockchain network. 
Blocks are continuously added to the chain to form an immutable record of all past transactions, creating what is known as the \textit{mine} of blocks~\cite{mineblocks}. 
The chosen block is determined by \textit{consensus protocols}~\cite{consensusprotocols} such as PoW~\cite{powinblockchain} or PoS~\cite{posinblockchain}.
Bitcoin~\cite{Bitcoinweb} is the first digital token that is implemented by blockchain technology. Ethereum first introduced the concept of smart contracts~\cite{smartcontract} in 2014, which are sets of codes that run on the blockchain.
Transactions can be initiated between accounts or even contracts to achieve complex functionalities,  such as transferring funds to another account or conducting calculations.

Smart contracts enable the creation of various digital tokens, extending beyond Bitcoin, which includes both fungible and non-fungible tokens (NFTs). Typically, these digital tokens are developed following specific protocols, such as the \textit{Ethereum Request for Comments} (ERC) in the case of Ethereum. In Ethereum, fungible tokens are established under the ERC-20 standard, while NFTs are primarily implemented through ERC-721 and ERC-1155 standards.

Specifically, ERC-721~\cite{erc721} is a standard protocol designed for NFTs where each token is bound to a single piece of digital content, e.g., art, music and video.
\textit{ERC-721 tokens cannot be divided into smaller units and must be traded as whole entities. }
However, it has limitations, e.g., difficulties in representing repeated artwork, and bulk transfers of a large number of tokens at once.
To address these, ERC-1155~\cite{ERC1155} was introduced, which allows multiple tokens to be associated with specific content, allowing all token holders to share access to that particular content.
It also facilitates multi-token operations that enable the transfer of multiple tokens in a single transaction.
Other standards, like Metaplex in Solona~\cite{solananft, solanaofficial} and BEP-721 in Binance~\cite{bep721, binanceofficial}, also support the nature of NFTs.

\subsection{Decentralized Application (DApp)}
Taking advantage of the interactions among smart contracts, developers can achieve complex functionalities in a \textit{decentralized application (DApp)}. DApps can operate on blockchain networks, ensuring data immutability, and guaranteeing the authenticity and reliability of information.
Moreover, after porting traditional finance applications on decentralized blockchains, \textit{decentralized finance (DeFi)} gradually emerges in 2017~\cite{defihistory}. DeFi relies on smart contracts and decentralized networks to offer services like lending, borrowing, trading, and earning interest on cryptocurrencies. 
Specifically, \textit{decentralized exchange (DEX)} is a special kind of DeFi DApp that implements cryptocurrency exchange.

To alleviate the burden of managing digital cryptocurrencies of users, wallets appear as browser plugins or standalone mobile applications~\cite{walletevalution}.
It assists users in initiating cryptocurrency transactions, conducting identity verification, and interacting with smart contracts. A hot wallet is a type of wallet that remains consistently connected to the internet, offering rapid access and transaction capabilities. However, due to its online nature, it is more vulnerable to cyber threats than its offline counterpart, known as cold wallets. The private key is a randomly generated unique alphanumeric character, utilized to prove ownership and control over digital assets stored in the wallet.

\begin{figure}[t]
    \centering
    \includegraphics[width=1\columnwidth]{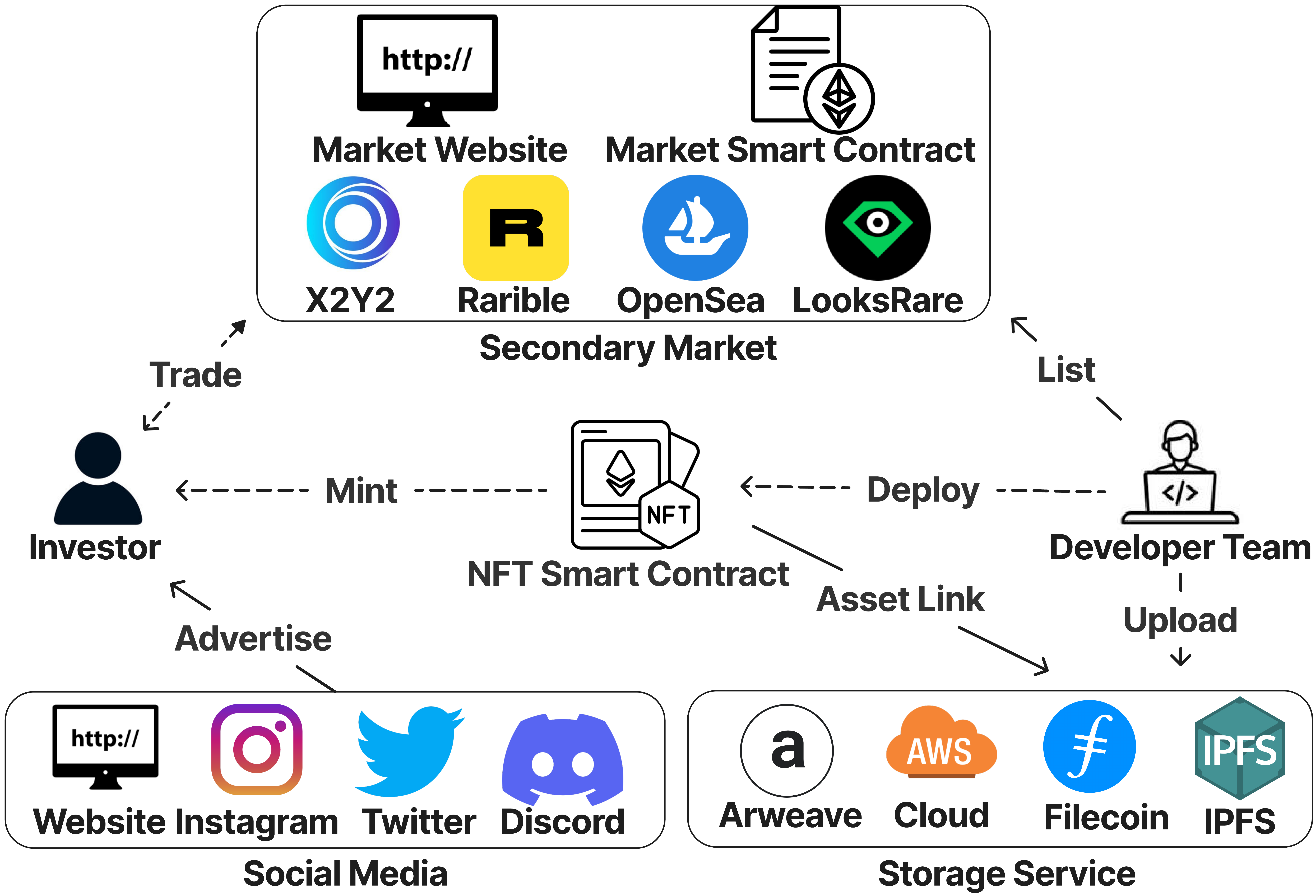}
    \vspace{-0.1in}
    \caption{The NFT life cycle shows the various stages of an NFT and the people taking part. Dashed lines stand for actions recorded on the blockchain, and solid lines represent off-chain activities not recorded on the chain. }
    \label{fig:Overview}
        \vspace{-0.1in}
\end{figure}

\subsection{NFT Life Cycle}
\label{sec:background:life:cycle}

Fig.~\ref{fig:Overview} shows the overall landscape of the NFT ecosystem. Generally, it can be divided into four phases, i.e., \textit{asset creation}, \textit{contract deployment}, \textit{NFT minting} and \textit{market circulation}.

\noindent
\textbf{Phase I: Asset Creation.}
The intrinsic value of an NFT primarily derives from its association with unique digital assets.
To create an NFT project, the developer first creates a collection of digital assets, e.g., a set of images, and upload them to \textit{off-chain} storage services, where users can access these digital assets via URIs (Unique Resource Identifiers).
These developers would advertise their projects on public social media platforms (like Twitter~\cite{twitterofficial}, Discord~\cite{discord}, and Instagram~\cite{instagramofficial}) and also their official websites.
Moreover, to consistently attract investors, they may also adopt some marketing strategies, like sharing project roadmaps or highlighting the artistic and economic potential of the project.

\noindent
\textbf{Phase II: Contract Deployment.}
The deployment of contracts is essentially the process of tokenizing digital assets, completing the mapping of off-chain assets to on-chain tokens. The developer deploys
the smart contract under a specific non-fungible token standard, e.g., the ERC-721 standard on Ethereum, linking the URI on-chain to the digital asset on the storage service.

\noindent
\textbf{Phase III: NFT Minting.}
NFTs are created by a special event called \textit{Mint}.
Two methods are commonly used to create NFT tokens for investors. 
First, investors can pay the mint fees to the smart contracts, exchanging the right to mint the token. 
Alternatively, the contract owner sometimes mints the tokens themselves and airdrops them to participants in the NFT ecosystem, freely distributing tokens and making them flow into the market.
Each specific NFT is unique once minted by a smart contract, serving as the identification for that particular token and its associated content.

\noindent
\textbf{Phase IV: Market Circulation.}
Participants in the NFT ecosystem typically perform trades of NFTs in secondary markets, i.e., the special decentralized applications focused on token trading, such as OpenSea~\cite{openseaofficial}, LooksRare~\cite{looksrareofficial}, and X2Y2~\cite{x2y2official}. 
After creating NFTs, developers list their tokens for sale on secondary markets~\cite{listnft}. 
Generally, these tokens are organized into collections~\cite {nftcollection} that refer to all tokens minted by the same smart contracts, which helps to gather all work in the same style, create a brand, and present it to the public. 
Secondary markets also provide an optional verification tag~\cite{verifyaccount} to help increase the credibility of the collections.
As decentralized applications, secondary markets have their smart contracts launched on the blockchain, such as Wyvern~\cite{wyvernofficial} or Seaport~\cite{seaportofficial} from OpenSea. These smart contracts implement different functions that help users trade in a secure environment. Buyers can make offers or bids before purchasing assets. Once accepted, the smart contract on the blockchain transfers the asset to the target account. 
During the process, the trade fees~\cite{nftmarketfee} are withdrawn by the secondary market, and creator fees~\cite{creatorfee} are withdrawn by the creator group.

\section{Reference Frame of NFT Security}
\label{sec:reference:frame}
We next present our methodology to construct a reference frame, which serves as the foundation for the analysis of security issues.

\subsection{Notations}
We first define the following notations used in the paper.

    \textbf{Security Incident.} 
    An NFT security incident refers to a \textit{concrete} case that \textit{can or may} result in an unexpected financial loss to one or multiple the following entities: (i) NFT holders, who own these tokens; (ii) NFT project teams, responsible for creating and managing NFT collections; and (iii) NFT sales platforms, which facilitate the trading and exchange of NFTs.
    Specifically, there are two types of incidents:
    (i) \textit{attacks} that exploit the vulnerabilities within smart contracts, or the defects within market and auxiliary service,
    and (ii) \textit{accidents} that do not explicitly involve proactive adversaries, but still causes the damage to relevant parties.
    For example, asset inaccessibility (see \S\ref{sec:assertinaccess}) is labelled as accidents. This is because even if attackers or scammers do not intentionally take the corresponding actions, they can still result in significant harm to the affected individuals in the community.

    \textbf{Security Issue \& Sub-issue.}
    A security issue refers to an abstraction of a set of incidents that possess similar causes and processes.
    If necessary, an issue can be further divided into sub-issues. For example, the reentrancy security issue consists of two sub-issues, i.e., single-function and cross-function (see \S\ref{sec:Reentrancy}). Similarly, each sub-issue is still composed of several incidents with similar root causes.
    Thus, security issue, sub-issue and incident form a multi-level inclusion relationship.

    \textbf{Category.} 
    Category refers to the abstract representation for the taxonomy and description of a group of related security issues.

    \textbf{Layer.}
    A layer is composed of components in the NFT ecosystem, and refers to an abstract location where a set of security issues occur.

\subsection{Data Collection}
\label{sec:data:collection}

\noindent
\textbf{Scope.}
To conduct a comprehensive analysis, we consider two types of materials as sources, i.e., \textit{academic papers} and \textit{security reports}.
Specifically, academic papers generally conduct systematic research on security issues, which allows them to go beyond reported security incidents and propose more general detection methods. Even further, they sometimes suggest defense mechanisms.
However, publishing academic papers often requires a significant amount of time for research, which means that academic papers on security issues tend to lag behind industry technical reports.
Thus, we also collect security reports published by flagship blockchain security companies because of their timeliness and professionalism. 
The gathering process continued till October 1st, 2023.

\noindent
\textbf{Approach.}
Our data collection for academic papers and security reports is divided into four steps.
First, we select a set of target sources. for security reports, we choose several reputable sources, i.e., \textit{Beosin~\cite{beosinteam}, SlowMist~\cite{slowmistteam}, Blocksec~\cite{blocksecteam}, Certik~\cite{certicteam}, DeFiHackLabs~\cite{defihacklab}, and De.Fi~\cite{defirekt}}.
For academic papers, we performed a thorough search on academic databases, i.e., \textit{Google Scholar, IEEE, ACM}, with the keywords: \textit{non-fungible token, NFT, blockchain security, DeFi}.
Second, we collect all related materials from these sources.
Then, we employ a rigorous review process. We engage two Ph.D. students with expertise in blockchain and security domains to meticulously review the collected materials. To ensure accuracy and eliminate potential biases, a paper is retained for our study only if both reviewers independently recognize its relevance to NFT security incidents or issues.
In the final step, we recognized that relying solely on our initial search terms could lead to overlooking relevant papers. 
Thus, we take all identified security issues as keywords to perform a comprehensive search on these databases again, like specifically searching \textit{reentrancy}.

Note that following the principle of selecting high-quality papers of previous SoK of blockchain~\cite{zhou2023sok}, we excluded papers published on \textit{arXiv} from our data collection approach.
We believe that since these papers do not undergo a peer-review process, some of them might be of lower quality, and excluding them could prevent fro introducing false positives into our dataset.

\noindent
\textbf{Dataset Overview.} 
Consequently, we have identified 213 security reports and 18 academic papers.
We noticed that certain incidents may be reported multiple times. 
After manually removing duplicates, 142 incidents from both security reports and academic papers are remained.
The dataset is open-sourced at \href{https://nftsok.github.io/}{link}.

\begin{table}[t]
\centering
\caption{The table shows all 16 kinds of security issues within the \textit{Contract Layer}, \textit{Market Layer}, and \textit{Auxiliary Service Layer}, separately. 
The number inside the parentheses indicates the quantity of collected incidents related to a specific security issue.
Note that, there are no incidents related to \textit{Asset-related Issue}. 
This is because both its sub-issues were identified solely from academic papers, and there is no technical report discussing them.
We use different symbols to mark the status of current research on IEI (Identifying Existing Incidents) and BP (Best Practice) topics. In particular, ``\CIRCLE'' indicates that a topic has been studied before, and ``\Circle'' means that it has not been technically explored in previous research.}
\label{tab:securityframe}
\resizebox{\columnwidth}{!}{%
\begin{tabular}{@{}llllll@{}}
\toprule
\textbf{Layer}                            & \textbf{Category}                            & \textbf{Security Issue}                     & \textbf{Sub-issue}               & \textbf{IEI} & \textbf{BP} \\ \hline
\multirow{9}{*}{\rotatebox{90}{Contract Layer}}           & \multirow{3}{*}{Unsafe External Call (6)}        & \multirow{2}{*}{Reentrancy}                 & Single-function Reentrancy       & \multicolumn{1}{c}{\CIRCLE}     & \multicolumn{1}{c}{\Circle}     \\
                                          &                                              &                                             & Cross-function Reentrancy        & \multicolumn{1}{c}{\Circle}      & \multicolumn{1}{c}{\Circle}     \\
                                          &                                              & DoS (Denial of Service)                                         & \multicolumn{1}{c}{-}            & \multicolumn{1}{c}{\CIRCLE}     & \multicolumn{1}{c}{\CIRCLE}    \\ \cline{2-6} 
                                          & \multirow{2}{*}{Insufficient Access Control (3)} & Ownership Verification Issue                & \multicolumn{1}{c}{-}            & \multicolumn{1}{c}{\Circle}      & \multicolumn{1}{c}{\Circle}     \\
                                          &                                              & Function Visibility Issue                   & \multicolumn{1}{c}{-}            & \multicolumn{1}{c}{\Circle}      & \multicolumn{1}{c}{\Circle}     \\ \cline{2-6} 
                                          & \multirow{3}{*}{Improper Business Logic (6)}     & \multirow{2}{*}{Conditional Expressing Bug} & Lack of Conditional Expression   & \multicolumn{1}{c}{\Circle}      & \multicolumn{1}{c}{\Circle}     \\
                                          &                                              &                                             & Erroneous Conditional Expression & \multicolumn{1}{c}{\CIRCLE}     & \multicolumn{1}{c}{\Circle}     \\
                                          &                                              & Variable-related Bug                        & \multicolumn{1}{c}{-}            & \multicolumn{1}{c}{\Circle}      & \multicolumn{1}{c}{\Circle}     \\ \cline{2-6} 
                                          & Bad Randomness (1)                               & Bad Randomness                              & \multicolumn{1}{c}{-}            & \multicolumn{1}{c}{\CIRCLE}     & \multicolumn{1}{c}{\Circle}     \\ \hline
\multirow{7}{*}{\rotatebox{90}{Market Layer}}             & Wash Trading (2)                               & Wash Trading                                & \multicolumn{1}{c}{-}            & \multicolumn{1}{c}{\CIRCLE}     & \multicolumn{1}{c}{\Circle}     \\ \cline{2-6} 
                                          & \multirow{2}{*}{Arbitrage (3)}                   & \multirow{2}{*}{Arbitrage}                  & Collection Offer Arbitrage       & \multicolumn{1}{c}{\Circle}      & \multicolumn{1}{c}{\Circle}     \\
                                          &                                              &                                             & Flashloan-based Arbitrage        & \multicolumn{1}{c}{\Circle}      & \multicolumn{1}{c}{\Circle}     \\ \cline{2-6} 
                                          & Rug Pull (45)                                     & Rug Pull                                    & \multicolumn{1}{c}{-}            & \multicolumn{1}{c}{\Circle}     & \multicolumn{1}{c}{\Circle}    \\ \cline{2-6} 
                                          & \multirow{3}{*}{Copyright Theft (3)}             & Sleepmint                                   & \multicolumn{1}{c}{-}            & \multicolumn{1}{c}{\CIRCLE}     & \multicolumn{1}{c}{\CIRCLE}    \\
                                          &                                              & \multirow{2}{*}{Counterfeit NFT}            & Name Counterfeit                 & \multicolumn{1}{c}{\CIRCLE}     & \multicolumn{1}{c}{\Circle}     \\
                                          &                                              &                                             & Image Counterfeit                & \multicolumn{1}{c}{\CIRCLE}     & \multicolumn{1}{c}{\Circle}     \\ \hline
\multirow{10}{*}{\rotatebox{90}{Auxiliary Service Layer}} & \multirow{3}{*}{Website Exploitation (10)}        & \multirow{3}{*}{Website Exploitation}       & Malicious Script Upload          & \multicolumn{1}{c}{\Circle}      & \multicolumn{1}{c}{\Circle}     \\
                                          &                                              &                                             & API Exploitation            & \multicolumn{1}{c}{\Circle}      & \multicolumn{1}{c}{\Circle}     \\
                                          &                                              &                                             & UI-based Fraud                   & \multicolumn{1}{c}{\Circle}      & \multicolumn{1}{c}{\Circle}     \\ \cline{2-6} 
                                          & \multirow{4}{*}{NFT-themed Phishing (52)}         & \multirow{4}{*}{NFT-themed Phishing}        & Website Phishing                 & \multicolumn{1}{c}{\CIRCLE}     & \multicolumn{1}{c}{\CIRCLE}    \\
                                          &                                              &                                             & Phishing Trojan                  & \multicolumn{1}{c}{\Circle}      & \multicolumn{1}{c}{\Circle}     \\
                                          &                                              &                                             & Spam NFTs                        & \multicolumn{1}{c}{\CIRCLE}     & \multicolumn{1}{c}{\Circle}     \\
                                          &                                              &                                             & Social Media Hacked              & \multicolumn{1}{c}{\Circle}      & \multicolumn{1}{c}{\Circle}     \\ \cline{2-6} 
                                          & \multirow{2}{*}{Asset-related Issue (0)}         & \multirow{2}{*}{Asset-related Issue}        & Asset Inaccessibility            & \multicolumn{1}{c}{\CIRCLE}     & \multicolumn{1}{c}{\Circle}     \\
                                          &                                              &                                             & Metadata Tampering               & \multicolumn{1}{c}{\CIRCLE}     & \multicolumn{1}{c}{\Circle}     \\ \cline{2-6} 
                                          & Private Key Compromised (11)                      & Private Key Compromised                     & \multicolumn{1}{c}{-}            & \multicolumn{1}{c}{\Circle}      & \multicolumn{1}{c}{\Circle}     \\ \hline
\end{tabular}%
}
\end{table}

\subsection{NFT Ecosystem Layering}
\label{sec:framework}

Building on the life cycle of NFTs depicted in \S\ref{sec:background:life:cycle} and widely-adopted frames of blockchain security from existing work~\cite{zhou2023sok}, we define three distinct layers for the whole NFT ecosystem, i.e., \textit{contract layer}, \textit{market layer}, and \textit{auxiliary service layer}.
We classify all security issues according to the specific location where they occur.
Specifically, the \textit{contract layer} comprises all security issues in contracts, resulting from code weaknesses that attackers could exploit. Note that, we focus on not only Ethereum blockchain, but also other blockchain platforms that can mint NFTs, like Solana.
The \textit{market layer} is composed of security issues happened on NFT markets. This involves examining unusual and harmful trades, as well as malicious actions affecting the sellers and buyers, or even everyone involved in the market. 
The \textit{auxiliary service layer} is defined as infrastructures that help in the regular workflow of NFT, such as social media or digital wallets. Therefore, we classify any security issues related to these infrastructures in this layer.

Table~\ref{tab:securityframe} presents all 16 security issues that we concern, as well as the category and layer they belong to. For each security issue, we further summarize the types of existing exploitation tactics. Moreover, we also conduct a comprehensive survey on current methods that can identify existing incidents and the best practices that can be adopted for each of security issue.
Based on these material, we will discuss the answers to the following questions:
\begin{enumerate}
	\item What constitutes a security incident for a security issue/sub-issue?  
	\item How to identify existing security incidents targeting a specific security issue/sub-issue?
	\item Which best practices can be adopted to avoid losses?
\end{enumerate}

\section{CONTRACT LAYER}
\label{sec:contract:layer}
We have summarized 7 kinds of security issues in contract layer.

\subsection{Unsafe External Call}
\label{sec:unfase:external:call}

In the lifecycle of a smart contract, it may invoke functions deployed by other accounts or even contracts. 
Since these functions are out of developers control, they may perform malicious operations, leading to unsafe external call.

\subsubsection{\textbf{Reentrancy}}
\label{sec:Reentrancy}
Reentrancy allows attackers to make multiple calls and callbacks to gain additional profits, which goes against the original intent of the function. 
According to the number of involved functions during the attack, two types of reentrancy have been observed in the NFT ecosystem.

\begin{lstlisting}[
caption={Case of single-function reentrancy: HyberBear.}, 
    label={list:hyberbearreentrancy}, 
    style=solidity
]
function mintNFT(...) public payable {
    for(uint i = 0; i < _numOfTokens; i++) {
        _safeMint(msg.sender, totalSupply() + 1);
    }}
function _safeMint(..) internal virtual {
    require(
    _checkOnERC721Received(address(0), to, tokenId, _data),
    "ERC721: transfer to non ERC721Receiver implementer" );
}
\end{lstlisting}

\noindent
\textbf{Sub-issue I: Single-function Reentrancy.} 
Single-function reentrancy refers to both the caller of an external call and the callee of the corresponding callback being the same function. Internal functions are used as functional modules of public functions, and they are considered as the same function. 
For example, the \texttt{mintNFT} function of HyberBear has been exploited~\cite{hyperbearreentranyc}, as shown in Listing~\ref{list:hyberbearreentrancy}.
After minting a token, \texttt{\_checkOnERC721Received()} (L7\footnote{L7 refers to the 7th line, we adopt this notation in the following.}) would call the designated function, where attackers can reinvoke the \texttt{mintNFT} function of HyberBear to repeat the mint process to gain profits. 

\begin{lstlisting}[
    caption={Case of cross-function reentrancy: OMNI.}, 
    label={list:omnireentrancy}, 
    style=solidity
]
function executeWithdrawERC721(...) external returns (uint256) {
    bool withdrwingAllCollateral = INToken(reserveCache.xTokenAddress).burn(...);
    if (userConfig.isBorrowingAny())
...
}
function burn(...) external virtual override onlyPool returns (bool) {
    IERC721(_underlyingAsset).safeTransferFrom 
}
function executeERC721LiquidationCall(...) external {
    userConfig.setBorrowing(
        liquidationAssetReserve.id, false);
}
\end{lstlisting}

\noindent
\textbf{Sub-issue II: Cross-function Reentrancy.}
Rather than entering the identical function, attackers may enter a different function that relies on the same states to achieve the reentrancy attack. 
OMNI is exploitable to this vulnerability~\cite{OMNIDegis,OMNIslowmist,blocksecomni}, and its implementation is shown in Listing~\ref{list:omnireentrancy}.
The \texttt{executeWithdrawERC721} function would external call the \texttt{burn} function of  (L6) to burn NFTs used as collateral tokens, subsequently returning the pledged assets to the user (L1). However, the \texttt{burn} function involves a callback mechanism (L7), thus providing an opportunity for attackers. During the callback process, attackers enter the different function \texttt{executeERC721LiquidationCall} (L9). The shared state variable indicating borrowing status is modified to \texttt{False} (L10).
Finally, the flow returns to the initial \texttt{executeWithdrawERC721} function, bypassing the check (L3), the attack is successfully executed. The attacker does not need to repay the loan.

\noindent
\textbf{Identifying Existing Incidents.}
The key characteristic of exploitation on reentrancy vulnerability is the presence of a \textit{circular call} in the traces. Thus, we can identify existing attacks by identifying circular call in transaction traces, where functions are nodes and invocations are directed edges.  
Moreover, to enhance the accuracy, additional rules can be incorporated. 
For instance, it is possible to check whether the initiator of the circular call makes profits or if there is a decrease in the \textit{Total Value Locked (TVL)} of the contract, which is a vital metric to show the value of one project.

\noindent
\textbf{Best Practice.}
The fundamental cause of reentrancy is the untimely update of state variables. The special of NFT contracts lies in their implicit calls to the \texttt{onERC721Received} and \texttt{onERC1155Received} functions. A contract must implement one of these to receive an NFT, helping prevent the NFT from being stuck in a contract that cannot handle NFTs. 
Developers should pay attention to this mechanism and utilize the community-endorsed Checks-Effects-Interactions pattern~\cite{checkseffectsinteractions}, which advocates that calling external contracts should always be the final step.

\subsubsection{\textbf{DoS (Denial of Service)}}
\label{sec:dos}
Attackers can take advantage of external calls and deliberately make external calls fail, causing the contract function to never be executed successfully, directly leading to a denial of service.
As shown in Listing~\ref{list:gas}, it is a DoS vulnerability discovered in Akutar~\cite{blocksecakutar}.
In the \texttt{processRefunds()} function (L1), there is a loop (L2 to L10) that refunds all users who have previously placed a bid, which consequently makes an external call to each of them (L6).
Since the external call lacks the check of exception, the malicious contract, deployed by attackers, may hide among the users. 
The malicious contracts can use a \texttt{revert} statement within their fallback or receive functions to ensure that refund transactions always fail, leaving no one able to get their fund back.

\noindent
\textbf{Identifying Existing Incidents.} 
DoS attacks always lead to failed transactions.
Thus, the information of failed transactions can assist in identifying existing DoS attacks, like the address of callers and the reason for the failure.
Specifically, if an external call fails due to a revert or the out-of-gas issue, it indicates a potential DoS attack.

\noindent
\textbf{Best Practice.}
This issue is not exclusive to NFT contracts. Any external call lacking proper exception handling could be vulnerable to be DoSed. 
NFT contract developers should avoid conducting multiple external calls in a single transaction, especially when calls are executed as part of a loop~\cite{dosbp}.
It is essential to always consider the possibility of external call failures and implement strategies to handle them.
There are existing work to help developers discover potential DoS in contracts.
For example, Ghaleb~\etal~\cite{ghaleb2022etainter} defines the target address of a call executed in the loop’s body as a sink to find data flows in which the taint source \textit{msg.sender} reached the sink.
Moreover, Grech~\etal~\cite{grech2018madmax} constructs intermediate representation based on the Contract EVM bytecode and uses Datalog-based inference rules for extracting contract properties to detect DoS vulnerability, while Liao~\etal~\cite{liao2023smartstate} uses smart contract bytecode and transaction trace to exact fine-grained state-dependency to detect DoS vulnerability.

\begin{lstlisting}[
caption={Case of transaction reversion call: Akutar.}, 
    label={list:gas}, 
    style=solidity
]
function processRefunds() external {
for (uint256 i=_refundProgress; gasUsed < 5000000 && i < _bidIndex; i++) {
...   
    bids memory bidData = allBids[i];
    if (refund > 0) {
        (bool sent,) = bidData.bidder.call{value: refund}("");
        require(sent, "Failed to refund bidder");
    }
...
}}
\end{lstlisting}

\subsection{Insufficient Access Control}
\label{sec:insufficient:access:control}

Access control addresses the question of \textit{who is allowed to do this}, defining who has the power to perform certain actions (e.g., minting new NFTs or withdrawing the funding from a contract) within a contract~\cite{accesscontrolopenzeppelin}.

\subsubsection{\textbf{Insufficient Ownership Verification}}
\label{sec:contract:ownership:verification}

\begin{lstlisting}[
caption={Case of owner verified issue: FlippazOne.}, 
    label={list:FlippazOneverifiedissue}, 
    style=solidity
]
function ownerWithdrawAllTo(address toAddress) public  {
    (bool success, ) = toAddress.call{value: address(this).balance}("");
    require(success, "Failed to withdraw funds.");
}
\end{lstlisting}

The contract needs to validate the address of the caller or the incoming address before executing privileged functions.
For example, the token transfer function can only be invoked by the token's owner himself or others who are authorized.
However, some contracts overlook the importance of confirming the owner before such sensitive operations.
The withdraw function in a smart contract is one of the privileged functions that requires validation of the caller. Lacking the verification on the caller's carrying authority can lead to financial loss.
One notable example is in the \texttt{ownerWithdrawAllTo()} function in FlippazOne, as shown in Listing~\ref{list:FlippazOneverifiedissue}.
The function here utilizes the low-level call in Solidity for fund transfers but lacks verification of the incoming recipient address (\texttt{toAddress}) before transferring the entire balance of the contract (\texttt{address(this).balance}) (L2). 
The vulnerability allows anyone to withdraw the entire balance within the contract.

\noindent
\textbf{Identifying Existing Incidents.}
When dealing with privileged functions, like token transfers and contract withdrawals, ownership verification is necessary. By actively monitoring transactions and comparing the addresses calling these functions with the actual owner addresses, potential insufficient validation issues can be identified.
This process can be assisted by the interfaces defined in NFT protocols, such as \texttt{ownerOf(uint256 \_tokenId)} from the ERC721 standard interface to query token owners and \texttt{owner()} to return the contract owner. 

\noindent
\textbf{Best Practice.} 
We can conclude that the essence of this vulnerability is the absence of validation on \texttt{msg.sender} and incoming addresses before critical operations.
Methods raised by existing studies can be directly adopted.
For example, Yang~\etal~\cite{yang2023definition} determined which functions required ownership verification by reading the ERC721 standard description. 
After that, they utilized symbolic execution to detect these functions without ownership verification.

\subsubsection{\textbf{Function Visibility Issue}}
\label{sec:funcitonvisibility}
The visibility of functions in Solidity determines the access level. Currently, there are four distinct levels, i.e., external (accessible from other contracts or accounts), internal (accessible from the contract itself and its derived contracts), public (accessible from all, i.e., combining external and internal), and private (accessible only from within the contract).
Improperly applying visibility for a function may lead to security issues.
A typical example is The Sandbox~\cite{Sandbox}, a famous project of NFT-based game. 
As shown in Listing~\ref{list:Sandbox_burn}, the contract have one external function named \texttt{burn()} (L6)  and another public one named \texttt{\_burn} (L1). The visibility of \texttt{\_burn} (L1) should have been set as \texttt{internal} and only can be called by \texttt{burn}, \texttt{burn} function can restrict parameters passed into \texttt{\_burn} to finish verification.
However, it is set incorrectly to \texttt{public}, anyone can call this function. Although there is a check statement between \texttt{from} and \texttt{owner} (L2), the two parameters can be constructed by any caller.

\begin{lstlisting}[
caption={Case of function visibility issue: The Sandbox.}, 
    label={list:Sandbox_burn}, 
    style=solidity
]
function _burn(address from, address owner, uint256 id) public {
    require(from == owner, "not owner");
    ...
    emit Transfer(from, address(0), id);
}
function burn(uint256 id) external {
    _burn(msg.sender, _ownerOf(id), id);
}
\end{lstlisting}

\noindent
\textbf{Identifying Existing Incidents.}
The primary purpose of internal functions is to encapsulate and reuse code for functions within the current contract.
Thus, the presence of internal calls initiated from external functions is a clear characteristic. Through using source code to construct the Abstract Syntax Tree (AST) to identify external function calls to the internal functions in the same contract. The AST contains information about the functions themselves, such as their names, and visibility modifiers, as well as control flow details like function call relationships, conditional branches, and more. Identify the functions being internally called, check their visibility, and discover potential visibility errors.

\noindent
\textbf{Best Practice.}
Both developers and users should pay attention to the visibility settings of functions, especially for internal functions.
For example, developers should follow standard function naming conventions~\cite{soliditystyle}, like naming internal functions beginning with an underscore. Following community-recognized development guidelines helps in avoiding errors and facilitates code audits, enabling the discovery of potential issues.

\subsection{Improper Business Logic}
\label{sec:improper:business:logic}

Security issues may arise due to mistakes made by programmers within the implementation of functions. 
According to our investigation, we categorize them into \textit{conditional expression bugs} and \textit{variable-related bugs}.

\subsubsection{\textbf{Conditional Expression Bug}}
\label{sec:conditionalexpression}
One of the critical purposes of conditional expressions is to protect critical functions from being arbitrarily invoked or limiting the number of certain variables. A buggy conditional expression always leads to unexpected behaviors.
According to our dataset, two types exist, i.e., \textit{lack of conditional expression}, and \textit{erroneous conditional expression}.

\noindent
\textbf{Sub-issue I: Lack of Conditional Expression. }
When conditional expressions are missed, some functions or statements that should only be executed a limited number of times will be called without restriction, which is likely to bring unexpected catastrophic consequences.
The well-known NFT collection \textit{The Bored Ape Yacht Club (BAYC)} serves as a classic example of lacking essential condition checks~\cite{unlimitmint}.
A limit on the number of NFTs that can be minted is crucial for ensuring the rarity and value of the NFT. 
Therefore, when minting a token, the contract needs to check whether the current supply is over the maximum supply of the collection.
However, as demonstrated in Listing~\ref{list:umlimitmint}, \textit{BAYC} does not have a necessary assessment of the total supply before minting one token (before L5), leading to the possibility of minting more tokens than the initially configured amount.

\begin{lstlisting}[
caption={Case of unlimited mint: BAYC.}, 
    label={list:umlimitmint}, 
    style=solidity
]
function reserveApes() public onlyOwner {        
    uint supply = totalSupply();
    uint i;
    for (i = 0; i < 30; i++) {
        _safeMint(msg.sender, supply + i);
    }
}
\end{lstlisting}

\noindent
\textbf{Sub-issue II: Erroneous Conditional Expression. }
In addition to a lack of conditional expression, there are instances where programmers might mistakenly write the conditional expression incorrectly. 
This can lead to significant issues, such as locking funds.
Erroneous conditional expressions leading to unintended behavior, such as locked funds, are exemplified by the Akutar project~\cite{blocksecakutar,beosinakutar}. As shown in Listing~\ref{list:lockingfunds}, the second requirement (L3) has a mistake, which should compare the \texttt{refundProgress} with \texttt{\_bidIndex} instead of \texttt{totalBids}. Consequently, the requirement will never be satisfied.
Over 11.5K Ether is locked in the contract forever since it can never withdraw the balance (L6).

\noindent
\textbf{Identifying Existing Incidents.}
Conditional expression can be regarded to the verification of state variables within a function. 
In the Ethereum Virtual Machine (EVM), the verification process involves accessing specific slots and comparing the retrieved values. If there is no access to storage slots containing critical variables, there might be a condition lack issue. 
Based on this characteristic, Yang~\etal~\cite{yang2023definition} used source code to establish a storage map and employed symbolic execution to detect the \texttt{mint} function that did not access the \texttt{totalsupply} storage slot, indicating a lack of \texttt{totalsupply} validation.
However, it is important to note that in NFT contracts, the \texttt{mint} function is a standard interface, with a roughly similar structure, making it easy to identify the critical variable \texttt{TotalSupply}.
On the other hand, in DeFi contracts utilizing NFTs, due to the diversity of the contracts themselves, identifying critical variables is not as straightforward. However, it's also possible to follow certain characteristics to discover potential critical variables, often associated with specific users (addresses) and stored within mappings.

\noindent
\textbf{Best Practice.}
Developers should pay attention to the setting of conditional expressions in contract development, including the specific placement and the conditions for comparison. When it comes to operations like transfers and withdrawals, particular attention should be given to their conditional checks.
Before the deployment, developers can apply some software analysis methods, like symbolic execution, to discover potential issues.

\begin{lstlisting}[
caption={Case of locking funds: Akutar.}, 
    label={list:lockingfunds}, 
    style=solidity
]
function claimProjectFunds() external onlyOwner {
    require(block.timestamp > expiresAt, "Auction still in progress");
    require(refundProgress >= totalBids, "Refunds not yet processed");
    require(akuNFTs.airdropProgress() >= totalBids, "Airdrop not complete");
    (bool sent, ) = project.call{value: address(this).balance}("");
    require(sent, "Failed to withdraw");        
}
\end{lstlisting}

\subsubsection{\textbf{Variable-related bug}}
\label{sec:variablebug}
The contract should include multiple variables for each transaction execution. 
However, using variables incorrectly could open up opportunities for attackers to exploit the contract.
The improper use of variables may lead to the funds locked in the contract forever. For example, 
there is one vulnerability discovered in CryptoPunks V1, which is the original CryptoPunks NFT after launching in 2017~\cite{withdrawfailureCrypto}. As shown in Listing~\ref{list:CrytoPunkV1}, \texttt{punkNoLongerForSale} (L2) will overwrite \texttt{offer.seller} for \texttt{msg.sender}, i.e.,  the buyer (L3). This leads to the actual seller not being able to withdraw money. This mistake almost led to the abortion of the project, but fortunately, the development team remedied it in time, which saved the most valuable NFT project today.

\noindent
\textbf{Identifying Existing Incidents.}
By collecting normal transactions from existing contracts and extracting operational characteristics of normal transactions, one can establish an expected transaction model. For instance, in NFT sales, there should be a decrease in the number of holding tokens of the seller and an increase of the one of the buyer.
By broadly establishing models for expected normal transactions, one can identify abnormal transactions that deviate from the expected pattern.
For example, in the above incident, where the account balance of the seller does not increase after selling an NFT, or the account balance of the buyer dose not decrease, it deviates from the expected pattern and indicates a potential attack.

\noindent
\textbf{Best Practice.}
The intrinsic property of variable-related issues is unexpected value modification. Thus, developers can adopt taint analysis to trace the sources that can modify variables to uncover unexpected data flows.
For instance, in the case of incident, by tracing the potential sources of for \texttt{Offer.seller}, one can uncover \texttt{msg.sender}, which indicates unexpected modification of values.

\begin{lstlisting}[
caption={Case of variable-related issue: CrytoPunk V1.}, 
    label={list:CrytoPunkV1}, 
    style=solidity
]
struct Offer {... address seller; ... }
function punkNoLongerForSale(uint punkIndex) {
    punksOfferedForSale[punkIndex] = Offer(false, punkIndex, msg.sender, 0, 0x0);
}
function buyPunk(uint punkIndex) payable {
    Offer offer = punksOfferedForSale[punkIndex];
    punkNoLongerForSale(punkIndex);
    pendingWithdrawals[offer.seller] += msg.value;
}
\end{lstlisting}

\subsection{Bad Randomness}
\label{sec:badrandomness}
Bad randomness in smart contracts refers to the random number is calculated solely based on determined seeds, which can be predicted by adversaries~\cite{qian2023demystifying}.
A buggy random number generating algorithm can potentially be exposed by attackers, creating opportunities for exploitation. 
Take Meebits as an example~\cite{meebitsaccident}, which is shown in Listing~\ref{list:badrandom}.
To mint a token, the \texttt{\_mint()} function must be called first (L1 to L6). 
This function in turn calls another function called \texttt{randomIndex()} (L7 to L11) to generate a random value, which is used as the token ID for a new NFT.
However, this randomness is generated based on a set of values, including blockchain data like nonce, difficulty, and timestamp (L9). All of these values are visible to anyone on the blockchain network.
Since Meebits NFTs have different rarities based on their IDs, their prices are highly sensitive to their rarities. Consequently, this vulnerability allowed malicious actors to mint NFTs with predictably high values.

\noindent
\textbf{Identifying Existing Incidents.} 
If an address consistently profits from contracts relying on random numbers, it may indicate flaws in the randomness, suggesting that the address might be exploited by this vulnerability.
From this characteristic, we can monitor the rarity of each NFT acquired by users through mint transactions. If an address frequently acquires rare NFTs through multiple transactions, there might be potential exploitation of randomness.

\noindent
\textbf{Best Practice.}
Bad randomness occurs when the random generator depends solely on determined factors, like block information (like block.blockhash) and environmental information (like msg.sender) in Ethereum.
Qian~\etal~\cite{qian2023demystifying} employ taint analysis techniques to track critical variables, such as \texttt{block.blockhash} and \texttt{block.coinbase}, thereby identifying potential bad randomness in smart contracts. 
Developers should guarantee that there are some unpredictable seeds, like the data returned by oracles.

\begin{lstlisting}[
caption={Case of bad randomness: Meebits.}, 
    label={list:badrandom}, 
    style=solidity
]
function _mint(address _to, uint createdVia) internal returns (uint) {
...
    uint id = randomIndex();
    emit Mint(id, _to, createdVia);
    emit Transfer(address(0), _to, id);
...}
function randomIndex() internal returns (uint) {
...
    uint index = uint(keccak256(abi.encodePacked(
        nonce, msg.sender, block.difficulty, block.timestamp))) % totalSize;
...}
\end{lstlisting}

\section{MARKET LAYER}
\label{sec:market:layer}
We next present five kinds of security issues related to NFT market.

\subsection{Wash Trading}
\label{sec:washtrading}
Wash trading refers to a type of deceptive practice where attackers trade tokens within a group, creating an illusion of prosperity with the intention of deceiving or scamming unsuspecting investors and users.
This type of activity has been deemed illegal in many countries~\cite{wikipediawashtrading}.
Several real-world washing trading cases have been uncovered by the media. For example, \textit{Meebits}, as one of the top collections in the NFT ecosystem, is charged to perform wash trading. 
The largest sale for tokens in that collection was only 3.95 Ether~\cite{meebitswashtrading}, before Jan 12th, 2022.
However, one seller sold \#13824 token of Meebits for \$49.5M~\cite{meebitswashtrade} this day to one buyer, which is ten million times than the largest sale before this day. 
Moreover, the token was again sold back from the previous buyer to the previous seller with nearly the same price.
This aligns with the definition of wash trading, where in a small group artificially generates a significant trading volume for an NFT project within a short period.
The huge amount of money between these two trades indicates that this is a wash trading.

\noindent
\textbf{Identifying Existing Incident.}
Previous studies~\cite{bonifazi2023performing, wen2023nftdisk, von2021nft, das2022understanding} have proposed several methods to detect wash tradings.
On the one hand, some of them adopt heuristic rule-based detection on NFT trades. For example, Von et al.~\cite{von2021nft} examined the top 52 NFT collections with the most impact on the secondary market and identified six different wash trading patterns, along with creating rules to detect them.
On the other hand, some have proposed a formal methodologies to find wash tradings in reducing false negatives to some extents.
For instance, Bonifazi et al.~\cite{bonifazi2023performing} built a mathematical model for NFTs by three defined parameters, to detect wash tradings.
However, despite numerous studies exist, none of them have specifically focused on the characteristics of attackers themselves, e.g., how many scam campaigns that perform wash trading are there in-the-wild.

\noindent
\textbf{Best Practice.}
It is necessary to defend against wash trading to eliminate fake prosperity in the NFT ecosystem. 
One possible approach is to promptly inform the community about projects involved in wash trading in real-time, with the goal of quickly alerting investors. 
The challenge lies in identifying wash trading shortly after it begins. 
Researchers can adopt some heuristics for detection, which can be assisted by the insights from prior instances of wash trading. Further, some advances techniques like AI-assisted detection can be explored.

\noindent
\subsection{Arbitrage}
\label{sec:arbitrage}
Considering the financial property, arbitrage exists within the NFT ecosystem.
Specifically, two types are observed, i.e., \textit{collection offer arbitrage}, and \textit{flashloan-based arbitrage}.

\noindent
\textbf{Sub-issue I: Collection Offer Arbitrage}
\label{sec:collection:offer:arbitrage}
A collection offer is like a \textit{wanted sign} for NFTs~\cite{nftcollectionoffer}, which is a feature offered by several trading markets, like Opensea, X2Y2, and LooksRare~\cite{x2y2collectionoffer,looksrarecollectionoffer}.
Specifically, NFT holders can sell an NFT to another account who has published a collection offer and obtain the attached money.
Thus, if a bot constantly monitors collection offers raised in NFT markets and NFT tradings, once there is an NFT being sold and an existing collection offer offers a higher value for that NFT, the bot will automatically buy it and immediately sell it to the collection offer.
For example, 
the transaction 0x4b5e\footnote{\href{https://etherscan.io/tx/0x4b5ea5668142d07a2b1d47fea6a04656284799d1c507f2dfc2722313509b144e}{https://etherscan.io/tx/0x4b5ea5...}}
is a collection offer arbitrage conducted by 0x9e93\footnote{\href{https://etherscan.io/address/0x9E9346E082D445f08FAB1758984A31648c89241A}{https://etherscan.io/address/0x9E9346...}}.
Specifically, 0x9e93 first bought the token from 0x4d91\footnote{\href{https://etherscan.io/address/0x4d91838268f6d6d4e590e8fd2a001cd91c32e7a4}{https://etherscan.io/address/0x4d9183...}}
by \$318.98 USD, then again sold it to 0x7d2d\footnote{\href{https://etherscan.io/address/0x7d2d6110ecc7e8bbfc276166cdaa74c41be1409b}{https://etherscan.io/address/0x7d2d61...}}
by \$456.04 USD in a single transaction on Opensea, which is in line with the format of collection offer arbitrage.

\noindent
\textbf{}

\noindent
\textbf{Sub-issue II: Flashloan-based Arbitrage}
Flashloan is a type of feature provided by some lending DApps, like Uniswap~\cite{uniswap}. 
By paying a relative small amount of fee, flashloan allows one account to borrow the a bunch of tokens, which have to be returned within a single transaction. 
Therefore, for individual investors, to execute a flashloan-based arbitrage, they first borrow the funds, then make a series of trades to profit from the price difference, and finally repay the flash loan with a small fee and return the borrowed assets. 
Some secondary marketplaces allow \textit{bidding} on NFTs (see \S\ref{sec:background:life:cycle}). 
Therefore, it is possible to bid one token at prices higher than the current market rates. 
As a result, some arbitrageurs seize the opportunity to profit from the price difference between the bid of a victim and the listed token, engaging in profitable arbitrage.
For example, Ape Coin recently engaged in a flashloan-based arbitrage~\cite{flashloan2}. Here's how it went down: first, the arbitrageur obtained a flashloan from a lending platform. Then, using the borrowed funds, the arbitrageur purchased an Ape Coin. 
Finally, the token was sold on LooksRare to the highest bidder, fetching a price six times higher than the original purchase.
Note that to ensure this transaction would be the first one in the next block, the arbitrageur transferred 38 Ether to F2Pool, a collaborative mining pool, encouraging block discovery and prioritizing transaction confirmation. Despite this significant cost, the arbitrageur still gained a substantial profit of 20 Ether in the end.

\noindent
\textbf{Identifying Existing Incidents.}
No published paper has measured the two kinds of arbitrage. 
First, for identifying existing incidents of collection offer arbitrage, researchers can focus on the key step, i.e., an account purchases and sells NFTs \textit{within a single transaction}.
If the price of purchasing is lower than the price of selling, we assume it is a collection offer arbitrage.
Second, the main difference between NFTs and regular flashloan-based arbitrage is the presence of NFTs. 
researchers can use a simple method to identify the existing incidents.
Specifically, we can monitor flashloan activities on lending platforms to see if borrowers are using these flashloans for tradings.

\noindent
\textbf{Best Practice.}
Note that, compared to other security issues, arbitrage is a legitimate practice of profiting from market rules. Therefore, its malicious intent is relatively minor.
One strategy to defend against such issues could involve front-running.
Specifically, defenders could identify the collection offer arbitrage opportunities before the arbitrageur does and swiftly execute the trade. 
Subsequently, they may return the assets, ensuring that the arbitrageur does not implement any profit.
To make this strategy effective, defenders need to swiftly identify opportunities and promptly execute transactions on the blockchain.
The research community has delved into how to make flashloan-based arbitrage bots faster than others~\cite{qin2021attacking}.

\noindent
\subsection{Rug Pull}
\label{sec:rugpull}
A rug pull is a set of behaviors, where scammers launch a project, attract investors' money, and then suddenly shut down the project or ghost investors while taking all invested assets. 
AniMoon~\cite{animoon} is a typical example.
First, the rug puller launched the project and advertised it through their Twitter and official website~\cite{animoonwebsite}. 
Second, after attracting enough investors and accumulating enough funds, the rug puller withdrew over \$6.3M worth of Ether from the smart contract of the project~\cite{animoonlost}, only 11 days after its launching. 
Third, they abandon the project permanently, by shutting down the Twitter account and stopping the update of its official website.
Now, the floor price of this project is low (less than 0.01 Ether)~\cite{animoonprice}, which indicates the unpopularity caused by the rug pull.
Note that, NFT-related DApps may also conduct a rug pull.
For example, SudoRare, a fork NFT exchange between two famous secondary markets, is reported as a rug pull~\cite{sudorarerugpull}.
Six hours after the launch of the platform, the team rugged the pull, stole the tokens from users and withdrew over \$820K worth of digital tokens\footnote{\href{https://app.zerion.io/0xbb42f789b39af41b796f6c28d4c4aa5ace389d8a/history}{https://app.zerion.io/0xbb42f7...}}.

\noindent
\textbf{Identifying Existing Incidents.}
Currently, there is no published work attempting to identify existing instances of rug pulls. 
Therefore, we propose a possible solution to address this gap.
NFT projects can be categorized into those with and without a presence on social media. 
According to the characteristic of rug pulls, where funds are quickly withdrawn after a rapid project launch, we can monitor the social media status of NFT projects, specifically, whether they are consistently updated.
Detecting rug pulls in NFT projects without social media is challenging. Therefore, we can employ heuristic observations and approach to identify rug pull events from multiple perspectives, developers have left their social media accounts not updated for an extended period, and etc.

\noindent
\textbf{Best Practice.}
Quickly assessing if a project is going to commit a rug pull in its early stage is crucial for defense against rug pulls. Therefore, it requires effectively utilizing limited and early-stage data of NFT projects.
Generally speaking, projects that tend to be rug pulled exhibit the following two characteristics in its early stage. Firstly, there is a substantial minting of tokens to earn mint fees. Secondly, they may display excessively high prices in the market. 
Therefore, detecting can be devised by focusing on these two aspects.

\noindent
\subsection{Copyright Theft}
\label{sec:copyright:theft}

Each NFT is linked to a piece of digital content.
Thus, scammers may steal copyright from famous digital assets, e.g., images, without permission, cheating the public and gaining illegal profit.
We divide this into two categories, i.e., sleepmint and counterfeiting NFTs.

\subsubsection{\textbf{Sleepmint}}
\label{sec:sleepmint}
Sleepmint refers to a kind of behavior, where the scammer exploited a backdoor in the contract, pretending to purchase currency from a famous artist, and subsequently sold it to the victim at a higher price.
Specifically, first, the attacker creates a smart contract with a hidden backdoor, allowing them to transfer the token from the owner to themselves without the owner's approval.
Then, the attacker initially mints a token associated with content to a celebrity and transfers it back from that celebrity via a backdoor. 
Since the blockchain records that the celebrity previously owned the token, the attacker then reclaims ownership.
Finally, the attacker claims that the NFT project was created by the celebrity, and they purchased the token from the celebrity. As a result, the attacker can sell it at a high price to a victim.
Beeple, a famous artist active in the blockchain community, had been once sleepminted~\cite{beeplesleepmint}. 
He was targeted by an anonymous person Personne, who minted an NFT\footnote{\href{https://etherscan.io/token/0x5FBbACf00ef20193a301a5BA20acf04765fb6DaC?a=40914}{https://etherscan.io/token/0x5FBbAC...}} (the token ID is 40914) to Beeple
, as the second edition of one Beeple's famous digital content, i.e., \textit{Everydays: The First 5,000 Days}.
After that, he drew it back and claimed that the token is created by Beeple.
Finally, he put it on sale on OpenSea~\cite{openseaofficial} and Rarible~\cite{raribleofficial} for sale. 
Fortunately, these two platforms both eventually deactivated the token and forbidded the sale.

\noindent
\textbf{Identifing Existing Incidents.}
According to the pattern discussed above, the key to identifying a sleepmint is to detect, for a token, whether there exists a transfer event conducted by an unapproved account after a mint event.
We note that Guidi et al.~\cite{guidi2022sleepminting} once investigated this issue, though their definition of sleepmint is overly broad that may lead to several false positives, which requires some manual effort to eliminate them.

\noindent
\textbf{Best Practice.}
Guidi et al.~\cite{guidi2023delving} also propose a method for preventing sleepmint by probing the actions of attackers, who are extracted from the result of their detection on sleepmint. 
Specifically, they leverage the existing incidents. 
They examine whether a transaction shows signs of connecting with an address linked to previous attackers involved in sleepminting. 
If such a connection is identified, the transaction should be blocked.

\subsubsection{\textbf{Counterfeiting NFTs}}
\label{sec:counterfeitnft}
Counterfeiting tokens in blockchain platforms is a widely spread malicious behavior. 
In the case of fungible tokens, counterfeiting can be generally conducted by imitating token names~\cite{xia2021trade}. When it comes to NFTs, except for name counterfeiting, the back-end digital assets can also be the target.

\noindent
\textbf{Sub-issue I: Name Counterfeit.} 
This type of counterfeit NFT usually has an identical or confusingly similar name to the victim NFT, exploiting the feature that blockchain does not place any name restrictions. 
For example, Bored Ape Yacht Club~\cite{BAYC} has an identical fake token\footnote{\href{https://etherscan.io/address/0x28ed3a5bef29d9f441700a49f7625fc081341d7a}{https://etherscan.io/address/0x28ed3a...}} which has been flagged by Etherscan. Fake NFT Moonbird\footnote{\href{https://etherscan.io/address/0x335cbebf70132e64eee4812b7160f6552662b2d6}{https://etherscan.io/address/0x335cbe...}} have a similar name to the official NFT Moonbirds~\cite{moonbirds}.

\noindent
\textbf{Sub-issue II: Image Counterfeit.} 
Image counterfeiting is the illicit act of stealing images from their original sources to generate NFTs. 
This practice commonly targets art images created by artists and other officially sanctioned NFT images. 
Using these artworks without authorization constitutes another form of creating counterfeit NFTs.
A typical example is~\textit{Starry Night Dog}~\cite{fakeminttweet}. The \textit{Starry Nights Dogs} is the most famous painting series of artist~\textit{ Aja Trier}, which was turned into an NFT without the authorization of the artist.

\noindent
\textbf{Identifying Existing Incidents.}
Das \etal ~\cite{das2022understanding} measured the existence of counterfeit NFTs in their paper, used the Levenshtein distance to measure the similarity of NFT names, and employed perceptual hashing algorithms to assess the similarity of NFT images.
However, it only focuses on the NFTs that have long names (i.e., over seven chars).
Therefore, they left some NFT collections with short names undetected, which introduces false negatives. For image comparison, only the similarity between images of different NFT collections was considered, disregarding comparisons with existing art images. Additionally, the perceptual hashing algorithm itself has a relatively slow execution speed.

\noindent
\textbf{Best Practice.}
For investors, to avoid being deceived by counterfeit NFTs, precautions should start from contract creation. Real-time monitoring of NFT contract creation allows for the identification of their NFT contract names. By comparing these names with official NFT names, potential name forgery can be detected. Identifying image forgery involves recognizing the URI where NFT assets are stored. Typically, this can be accomplished by calling the contract's \texttt{tokenURI} to retrieve metadata URIs, from which image URIs can be read. These URIs can then be used to fetch images for similarity comparison. Establishing a database comprising existing NFT images and artwork images facilitates conducting similarity comparisons for newly created NFT images.

\section{AUXILIARY SERVICE LAYER}
\label{sec:auxiliary:service}

This section focuses on the security issues located in the auxiliary service layer, like the security issues related to front-end of secondary marketplaces.

\subsection{Website Exploitation}
\label{sec:websiteexploit}
NFT-related DApps, while leveraging traditional web pages to offer user-friendly interfaces, also inherit the risks and vulnerabilities of conventional web2 environments, including threats like \textit{malicious script upload}, \textit{API exploitation}, and \textit{UI-based fraud}.

\noindent
\textbf{Sub-issue I: Malicious Script Upload.}
Attackers exploit vulnerability in the website to upload malicious scripts.
These scripts are able to alter NFT information (like URIs), redirect users to fraudulent sites, or even manipulate transactions.
A prime example is PREMINT~\cite{walletdrainer}, an NFT tool that helps NFT projects create access control conditions for NFT mints to prevent bots from minting the NFT supply. It provides a front-end visual interface to help users visualize the NFT mint settings. Primint front-end integrates boomerang~\cite{Boomerang}, which allows users to upload images to the back-end server. However, due to a vulnerability within the boomerang, attackers can bypass the verification rules and upload a malicious JavaScript file~\cite{PREMINTOfficialBriefing}.
Using malicious code, attackers can generate deceptive transactions. If victims unknowingly sign these, it gives the attackers the approval to transfer the victims' NFTs and Ether~\cite{AnalysisofPremint,preminthack}.

\noindent
\textbf{Sub-issue II: API Exploitation.}
The application programming interface (API) on NFT secondary marketplaces may be exploited to get sensitive information, like the price of a specific NFT.
Opensea NFT Marketplace once suffered this one with the exploiter withdrawing around 332 ETH~\cite{openseafrontened,analysisopenseafrontened,apiexploit} worth in 2022.
Take the biggest sale \textit{BAYC} as an example\footnote{\href{https://etherscan.io/token/0xbc4ca0eda7647a8ab7c2061c2e118a18a936f13d?a=8274}{https://etherscan.io/token/0xbc4ca0...}}.
First, to avoid huge canceling fees, the victim managed to transfer his NFTs to another wallet.
This action should have canceled the listings of NFTs and these NFTs should not appear in the frontend of OpenSea~\cite{openseaofficial}, however, they are still on listings on Rarible~\cite{raribleofficial} which gains the data through the API \textit{Get All Listings}~\cite{getalllistingsAPI}  of OpenSea.
The victim himself thought that the listings were canceled and transferred the NFTs back to the original wallet, which set the price as 0.
The opportunist rapidly bought the tokens from the former listings without any price and made a huge loss to the victim for about 332 Ether.

\begin{figure}[t]
	\centering
        \subfigure[Original one.]{
            \includegraphics[width=0.22\linewidth]{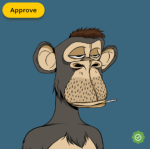}
            \label{fig:realswapscam}
        }
        \hfill
        \subfigure[Fake ones with green icon embedded.]{
            \includegraphics[width=0.69\linewidth]{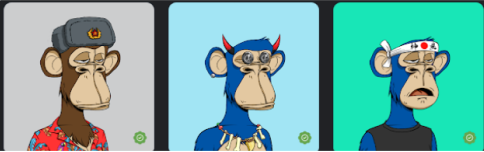}
            \label{fig:fakeswapscam}
        }
        \caption{Real BAYC \vs fake BAYC.}
\end{figure}

\noindent
\textbf{Sub-issue III: UI-based Fraud}
\label{sec:UI:fraud}
The design of the user interface (UI) could provide the possibility for the scam of victims.
Attackers have exploited the user interface of markets to commit fraud, which has been reported~\cite{NFTswapattack,NFTswapattack2}.
Typically, on Swap.Kiwi, an NFT secondary marketplace, an official and verified NFT will be labeled a green checkmark by the market, as Fig.~\ref{fig:realswapscam} shows.
However, a malicious user created a fake BAYC NFT with a green checkmark overlapped manually, as shown in Fig.~\ref{fig:fakeswapscam}.
The victim did not notice such a subtle difference.
Therefore, he swapped his real BAYC NFT with the fake one on Swap.Kiwi.
This caused nearly \$570K in economic losses to the victims.

\noindent
\textbf{Identifying Existing Incident.}
Targeting these traditional web2 vulnerabilities, there are some powerful ready-made scanners like Nessus~\cite{nessus}, Burp Suite~\cite{burpsuite}, etc., which can detect potential web vulnerabilities.

\noindent
\textbf{Best Practice.}
To prevent such vulnerabilities, we summarize best practices from different perspectives.
Standing from the developer side, they should conduct regular code reviews and static analysis, adhere to secure coding standards, and implement stringent input validation.
On the user side, they should verify the authenticity of the websites visited and avoid clicking on any links or downloading attachments from unknown sources.

\subsection{NFT-themed Phishing}
\label{sec:nft:themed:phishing}

Phishing is a deceptive practice where attackers pretend themselves as trustworthy entities to extract sensitive information from victims.
NFT-themed phishing includes website phishing, phishing Trojans disguised as NFT games, spam NFTs that use NFTs as phishing link carriers, and hacking NFT project social media to spread phishing links.
The following will provide a detailed overview. 

\subsubsection{\textbf{Website Phishing}}
\label{sec:websitephishing}
In website phishing, attackers use fraudulent websites that mimic the appearance and domain names of official sites to trick users into signing malicious transactions, thereby stealing their assets.
Website phishing is also practical in the NFT ecosystem.
The typical case is \textit{https://moonbirds-exclusive.com/}, which is targeting the famous project \textit{MoonBirds}~\cite{slowmistphising}. 
Upon entering the website, it will request users to connect their wallets.
When the user clicks the connect wallet button, a signature request warning pops out on MetaMask. This is the first malicious transaction, requesting users to sign using \texttt{eth\_sign}, which is the highest authority signature~\cite{metamaskdocs}. Coupled with the address of the signer, this signature data can be reused to sign any other transaction. The second malicious transaction is \texttt{SetApproveForAll} authorization, the authorized address can freely transfer the NFTs of the grantor~\cite{setApprovalForAll}. Attackers can steal the signatures of users and tokens through phishing.

\noindent
\textbf{Identifying Existing Incidents.}
DNSTwist~\cite{dnstwist} generates similar domains by squatting from an official domain to find potential phishing websites.

\noindent
\textbf{Best Practice.} 
The best practice to defend against website phishing attacks is to constantly monitor the creation of web pages and set alerts for any suspicious activities.
He \etal ~\cite{he2023txphish} sets up a cert-stream server to receive live updates from the Certificate Transparency Log network, thus retrieving suspicious domains, including domains generated by squatting in real-time.

\subsubsection{\textbf{Phishing Trojan}}
\label{sec:phishingtrojan}
Phishing trojans are designed as ordinary software, appearing genuine but actually containing malicious programs that steal crucial user information~\cite{trojanphishing}. 
NFT-themed phishing trojans have emerged within the ecosystem.
In the CthulhuWorldP2E NFT game project, the project's operator distributed malware masked as game beta files on their Discord channel~\cite{cthulhuworldp2e}. This malicious software was later identified as \textit{RedLine Stealer}. Note that, RedLine Stealer~\cite{RedLineStealer}, available since 2020~\cite{redlinemalware}, is frequently concealed within Phishing trojans. Its latest iteration is capable of extracting data from cryptocurrency wallets, especially those functioning as browser extensions, such as \textit{MetaMask}.

\noindent
\textbf{Identifying Existing Incident.}
Paying close attention to user reports and complaints regarding unusual software behavior or unauthorized access to information.

\noindent
\textbf{Best Practise.}
The best practices for combating Phishing Trojans should be approached from a user perspective, aiming to enhance user awareness and safety.

\subsubsection{\textbf{Spam NFTs}} 
\label{sec:spamnft}
Using a unique airdrop mechanism, NFTs containing phishing links in their description are mass-distributed to users, similar to sending spam emails~\cite{spamnft}. 
The description of the NFT contains language that lures users to click on phishing links.
The NFT collection \textit{Lets Walk} is typical spam NFT phishing~\cite{spamnftcase}. 
First, the tokens of the collections are airdropped to different users without their approval.
Second, there is a phishing link in the description of this NFT, which is a vehicle to spread phishing sites. 
These NFT descriptions contain enticing language to lure users to a phishing site for free mint.
Consequently, the phishing site lures the victim to sign a malicious transaction, which enables them to transfer all their digital assets.

\noindent
\textbf{Identifying Existing Incident.}
\textit{Forta}~\cite{forta} provides a heuristic rules-based approach~\cite{spamnftdetector} for detecting spam NFTs, considering factors such as false total supply and the scale of an airdrop.

\noindent
\textbf{Best Practice.} 
After understanding this security issue, we have summarized the best practice from various perspectives. 
Specifically, from the standpoint of users, they need to be careful about any sudden appearance of unfamiliar links in their accounts. 
If so, they should not be deceived by them. 
From the platform perspective, educating users about this situation and standardizing the descriptions of coins on their platform can be beneficial. 
From the viewpoint of researchers, deploying a monitoring system to oversee the descriptions of airdropped tokens is malicious.

\noindent
\subsubsection{\textbf{Social Media Hacked}}
\label{sec:social:media:hacked}
Recall that, some NFT collections constantly attract investors through social media~\cite{kapoor2022tweetboost}, such as \textit{Twitter}~\cite{twitter}, \textit{Discord}~\cite{discord}, \textit{Telegram}~\cite{telegram} (see \S\ref{sec:background:life:cycle}).
Attackers could hijack these social media to post fake news or distribute phishing links.
The most renowned NFT project, \textit{BAYC}, had its \textit{Discord} channel hacked~\cite{BAYC}, where hackers posted phishing links, stealing approximately \$360K worth of NFTs~\cite{BAYCDiscordhack1,BAYCDiscordhack2}.

\noindent
\textbf{Identifying Existing Incident.}
Whether NFT social media has been hijacked can ultimately only be confirmed by the account owner, hence it can only be assessed based on the official disclosure of information. 
Typically, after a social media hack, attackers tend to post numerous false messages often related to giveaways, including phishing links to lure users into clicking. 
By continuously monitoring official project social media channels, the discovery of phishing links suggests potential account hijacking.

\noindent
\textbf{Best Practise.} 
The best way for users to prevent this sub-issue is to target~\textit{social media hacked}  use strong and unique passwords, or multi-factor authentication. Moreover, they should be cautious with suspicious emails and links.

\subsection{Asset-related Issue}
\label{sec:asset-related:issue}
Recall that assets are stored in different storage services and are linked via URIs. 
However, the combination of on-chain data and off-chain assets raises some security concerns that could harm this connection.
Specifically, two kinds of security incidents are observed, i.e., \textit{asset inaccessibility} and \textit{metadata tempering}.

\noindent
\textbf{Sub-issue I: Asset Inaccessibility.}
\label{sec:assertinaccess}
The value of each NFT solely depends on its bound real-world digital asset, where the indexing is accomplished by URIs. In other words, if a URI is invalidated making the real-world asset inaccessible, the corresponding NFT will be valueless.

\noindent
\textbf{Sub-issue II: Metadata Tampering.}
\label{sec:metadata:tampering}
The metadata of a token contains the corresponding URI to the bound asset. 
Consequently, if the metadata is altered maliciously, the token loses its meaning. 
The ERC-721 standard for NFTs actually permits the modification of a token's metadata.

\noindent
\textbf{Identifying Existing Incidents.}
For the first sub-issue, current work~\cite{wang2023nfts} has tried to check the data accessibility from three stages, i.e, the \textit{TokenURI} to \textit{Metadata}, \textit{Metadata} to \textit{AssetURI}, and \textit{AssetURI} to \textit{Asset}.
Consequently, according to the statistics, 10 centralized platforms account for hosting 79.04\% of NFTs’ off-chain assets, indicating that NFTs are stored is a centralized way to some extent. 
However, among them, only 33.77\% of NFTs’ assets can be retrieved.
This result suggests that approximately two-thirds of NFT have already lost their value because of the asset inacessibility.
For the second sub-issue, Das et al.~\cite{das2022understanding} once monitored the metadata of all external assets on \textit{OpenSea} three times over a period of six months at regular intervals. They observed that 2.89\% of metadata and 1.15\% of assets had been changed between the first two and the last two crawls, respectively. 
Note that this method assumes all changes are malicious by default, which is a limitation. 
However,  there are cases where metadata changes are necessary because of the asset inaccessibility. 
Therefore, researchers should conduct further checks to determine if the original assets remain accessible to confirm whether such actions are indeed malicious.

\noindent
\textbf{Best Practice.}
Two approaches could be adopted to mitigate \textit{asset inaccessibility}.
First, researchers could tackle the problem of assets being inaccessible through the storage service. 
However, this is challenging because it requires us to modify the storage service protocol to ensure perpetual asset accessibility.
Second, researchers could concentrate on the URIs of NFTs. The NFT standard permits the modification of the URI of an NFT. 
However, this also serves as a significant vulnerability, since it's unclear whether the changes to NFTs are malicious or not, as elaborated in the next incident.
Moreover, at present, there are no widely recognized suggested defense methods for \textit{metadata tampering}.
But some alternative approaches exist.
One of them is to monitor the permissions for the relevant functions that are able to change the metadata in the contract, ensuring that only the owner or authorized entities can invoke it.
Another strategy involves implementing a dynamic monitoring system. This system checks whether the metadata of tokens in the market has been modified and whether their original assets are still accessible.

\subsection{Private Key Compromised}
\label{sec:private:key:compromised}
Recall that, a crypto wallet is a digital tool that manages blockchain-based assets (see \S\ref{sec:nft:background:primer}). 
Users utilize cryptographic public-private key pairs to access wallets and sign transactions. 
However, their private key or digital signature might be leaked, e.g., through a phishing site, which may result in loss of account funds.
LiveArtX~\cite{Liveart}, an NFT art platform, serves as a typical example. The compromise of its wallet private key led to unauthorized NFT transfers and incurred a loss of around \$39K.

\noindent
\textbf{Identifying Existing Incident.}
After a private key is compromised, it can result in an abnormal decrease in account funds, often transferred to previously unassociated accounts.

\noindent
\textbf{Best Practice.}
Through the use of hardware wallets as physical vessels to store private keys, securely enclosing private keys and preventing online exposure. 
Using a multi-signature wallet which mandates the presence of multiple keys for transaction approval to increase security. 
Elevating Phishing Awareness enlightens users about the perils of inadvertently divulging keys to deceitful schemes. By integrating these strategies, the potential risks linked to private key breaches can be significantly diminished.

\section{Discussion}
\subsection{Lessons Learned}
\label{sec:discussion}

\noindent
\textbf{Gap between Industry and Academy.}~\textit{Our systematic survey shows that there is a lack of a comprehensive understanding of NFT security issues in academic research.}
Current research on NFT security only addresses a limited portion of the overall NFT security issues. 
In particular, for all security issues,  the majority of them have not been discussed in published papers on NFT security. 
Therefore, there is a lack of a comprehensive understanding of NFT security issues in academic research, indicating a notable gap between academic and industry awareness in this field.

\noindent
\textbf{Detectable Security Issues.}
\textit{Many security issues, have not been extensively measured on a large scale in previously published articles.}
Note that some security issues are isolated and not widespread, e.g., UI-based fraud (see \S\ref{sec:UI:fraud}), making it challenging to measure them comprehensively on a large scale. 
However, we have identified security issues that are common and can be systematically identified or measured. 
Among these, academic research has explored issues like wash trading and asset inaccessibility. 
Nevertheless, many security issues, such as spam NFTs, have not been extensively measured on a large scale in previously published articles. These security issues that can be detected or measured carry significant academic value. We also offer initial detection strategies to inspire future research in this area in this paper.

\noindent
\textbf{Limited Understanding of Attackers.}
~\textit{Even if certain security issues have been claimed to be thoroughly studied, our understanding of attackers is still limited.}
When exploring NFT security issues in depth, despite with understanding of how to detect them and how common they are, most studies do not thoroughly study the attackers themselves. Take wash trading, for example. Many papers study the problem, but none of them go deep into understanding the attackers. Consequently, our knowledge of the attackers is limited, and we still have unanswered questions, like whether attackers work together in organized groups.

\noindent
\textbf{Absence of Defense System.}
~\textit{Apart from a few specific security issues, we have not seen researchers suggesting defense systems for most security problems.}
Considering the widespreadness, severity, and intractability nature of NFT security issues, it becomes clear that having a defense system is of paramount importance in tackling these security concerns. 
While it may not be possible to entirely eliminate some security issues, we can still implement methods, like providing valuable recommendations by automatic process, to reduce the risks associated with these problems.
However, except for a few specific security issues, such as sleepmint (even though we have identified its limitation, as discussed in \S\ref{sec:sleepmint}), we have not observed any researchers proposing defense systems for most security issues.

\subsection{Threats to Validity}
Some limitations exist in our work that may affect the validity.
First, we gather reports from renowned sources of technical reports and peer-reviewed papers. Consequently, we do not consider reports from smaller or emerging companies and communities, and all pre-print websites. We argue that we have made our best effort to collect data from the most dependable and available sources.
Though our data collection procedure is heavily based on a manual process, to establish the most reliable data set, each report or paper is meticulously reviewed by two students who have paid attention in blockchain for more than two years.
Second, we do not discuss the detection or defense methods for all security issues, as some of them are specific cases and not applicable in a general context. 
Nevertheless, we have either summarized existing research or outlined potential methods for addressing all general security issues.

\section{Related Work}

\noindent
\textbf{Cryptocurrency SoK.}
In 2019, Praitheeshan et al. conducted a survey on DeFi in the Ethereum network, where they identified 16 security concerns and provided some solutions~\cite{praitheeshan2019security}.
Saad et al. categorized 22 vulnerability factors into three areas: blockchain structure, P2P system, and blockchain applications, studying their relationships among miners, mining pools, users, and exchanges~\cite{saad2019exploring}.
In 2020, Chen et al. found 40 vulnerabilities, examined 29 real-world attacks, and discussed 51 defenses related to blockchain technology~\cite{chen2020survey}.
In 2021, Werner et al. conducted a comprehensive overview of DeFi, covering its basics and security issues~\cite{werner2022sok}.
In 2022, Zhou et al. conducted another comprehensive overview, with a focus on DeFi incidents and security. 
They analyzed 77 academic papers, 30 audit reports, and 181 real-world incidents~\cite{zhou2023sok}.
Our research is centered on NFTs, which represent a specific aspect of DeFi. 
Note, there are no SoK specifically addressing the security issues of NFTs. 
This lack of research has motivated our work, which is the first and aims to contribute to future research in this area.

\noindent
\textbf{NFT Security.}
In 2022, Das et al. provided an overview of security issues in the NFT ecosystem related to marketplaces, external entities, and user behaviors~\cite{das2022understanding}. 
Bhujel et al. discussed security concerns in NFT marketplaces, particularly regarding private keys~\cite{bhujel2022survey}. 
Von et al. identified six wash trading patterns and established rules for their detection~\cite{von2021nft}.
Kapoor et al. explored the influence of social media on NFT value~\cite{kapoor2022tweetboost}. 
Wang et al.~\cite{wang2023nfts} study the storage service of NFT off-chain assert and found the problem of asset inaccessibility.
However, these studies do not cover all NFT security issues, and our research highlights a significant gap between industry and academic perspectives on NFT security.

\section{CONCLUSION}
The non-fungible token is a key component in the Web3 ecosystem, that has attracted great from both academia and industry. In this paper,
we present the first SoK of NFT security, covering 13 major categories of NFT security issues, by scrutinizing 142 security incidents.
We have provided potential detection and defense methods for those security issues unsolved, and shown that NFT security issues are widespread, severe, and intractable.
We have identified the gap between industry and academia,  
which will boost the future research of NFT security.

\bibliographystyle{ACM-Reference-Format}
\bibliography{cite}

\end{document}